\documentclass[12pt,preprint,apj]{emulateapj}

\newcommand{\bfrsbo}{f_{\rm SB}}
\newcommand{\bfrsbw}{f_{\rm WB}}
\newcommand{\ha}{\rm{H}\alpha}
\newcommand{\hb}{\rm{H}\beta}

\newcommand{\eddr}{L_{\rm [OIII]}/M_{\rm BH}}

\newcommand{\oth}{{\rm [OIII]}~\lambda5007}
\newcommand{\ntw}{{\rm [NII]}~\lambda6584}

\newcommand{\mstar}{M_{\rm star}}

\shorttitle{Do bars trigger AGN activity?}
\shortauthors{Lee et al.}

\begin{document}

\title{Do bars trigger activity in galactic nuclei?}

\author{Gwang-Ho Lee$^{1}$, Jong-Hak Woo$^{1}$, Myung Gyoon Lee$^{1}$, Ho Seong Hwang$^{2,3}$, Jong Chul Lee$^{1,4}$, Jubee Sohn$^{1}$, and Jong Hwan Lee$^{1}$}
\affil{$^{1}$Department of Physics and Astronomy, Seoul National University , 1 Gwanak-ro, Gwanak-gu, Seoul 151-742, Republic of Korea}
\email{ghlee@astro.snu.ac.kr}

\email{woo@astro.snu.ac.kr}

\email{mglee@astro.snu.ac.kr}

\affil{$^{2}$CEA Saclay/Service d'Astrophysique, F-91191 Gif-sur-Yvette, France}
\affil{$^{3}$Smithsonian Astrophysical Observatory, 60 Garden Street, Cambridge, MA 02138, USA}
\email{hhwang@cfa.harvard.edu}

\affil{$^{4}$Korea Astronomy and Space Science Institute 776, Daedeokdae-ro, Yuseong-gu, Daejeon 305-348, Republic of Korea}
\email{jclee@kasi.re.kr}

\email{jbsohn@astro.snu.ac.kr}

\email{leejh@astro.snu.ac.kr}


\begin{abstract}

We investigate the connection between the presence of bars and AGN activity,
using a volume-limited sample of $\sim$9,000 late-type galaxies with axis ratio $b/a>0.6$
and $M_{r} < -19.5+5{\rm log}h$ at low redshift ($0.02\le z\lesssim 0.055$),
selected from Sloan Digital Sky Survey Data Release 7.
We find that the bar fraction in AGN-host galaxies (42.6\%) is $\sim$2.5 times
higher than in non-AGN galaxies (15.6\%), and that the AGN fraction is a factor
of two higher in strong-barred galaxies (34.5\%) than in non-barred galaxies (15.0\%).
However, these trends are simply caused by the fact that AGN-host galaxies are on average
more massive and redder than non-AGN galaxies because the fraction of strong-barred galaxies
($\bfrsbo$) increases with $u-r$ color and stellar velocity dispersion.
When $u-r$ color and velocity dispersion (or stellar mass) are fixed,
both the excess of $\bfrsbo$ in AGN-host galaxies and the enhanced AGN fraction
in strong-barred galaxies disappears.
Among AGN-host galaxies we find no strong difference of the Eddington ratio distributions
between barred and non-barred systems.
These results indicate that AGN activity is not dominated by the presence of bars,
and that AGN power is not enhanced by bars.
In conclusion we do not find a clear evidence that bars trigger AGN activity.

\end{abstract}


\section{Introduction}

Galactic bars are believed to play a crucial role in galaxy evolution.
By reducing angular momentum, galactic bars can efficiently transport
gas from outer disk to the central kiloparsec scale
\citep{lyndenbell79,sellwood81,albada81,combes85,pfenniger91,heller94,bournaud+02,athanassoula+03, jogee+06},
as demonstrated by a number of numerical simulations
\citep[e.g.,][]{roberts79, athanassoula92,friedli93,maciejewski+02,regan+04}.
The bar-driven gas can cause a mass accumulation within the Inner Lindblad Resonance (ILR),
leading to the destruction of bars and the formation of psuedo-bulges
\citep{hasan90,pfenniger90,hasan93,norman96,das+03,shen+04,athanassoula+05,bournaud+05}.
Numerous observational studies have found the characteristics of the bar-driven gas:
i.e., inflow velocities from CO emission \citep[e.g.,][]{quillen95,benedict96}
and from $\ha$ emission \citep[e.g.,][]{regan97}, higher $\ha$ luminosities in barred galaxies
than in non-barred galaxies \citep[e.g.,][]{ho97}, and higher molecular gas concentrations
in the central kiloparsec region of barred galaxies \citep[e.g.,][]{sakamoto99,sheth+05}.

Because of the high efficiency of gas inflow toward the central region of galaxies,
bars are often invoked as a trigger of nuclear star formation.
Enhanced nuclear star formation has been found in the central regions of barred spiral galaxies
\citep[e.g.,][]{heckman80,hawarden86,devereux87,arsenault89,huang96, ho97,martinet97,emsellem+01,knapen+02,jogee+05,hunt+08,Ann+09}.
Some statistical studies presented high bar fractions among star-forming galaxies: e.g., 61\% in \citet{ho97},
82\%-85\% in \citet{hunt99}, and 95\% in \citet{laurikainen+04}.

Bar-driven gas inflow has been also considered as a mechanism for triggering active galactic
nucleus (AGN) activity \citep{combes+03}. For the past three decades, much effort has been devoted to understand
the connection between the presence of bars and AGN activity.
However, it is not yet clear whether bars transport gas down to the vicinity of supermassive black holes (SMBHs).
Several observational studies claimed that the fraction of barred galaxies is higher in AGN-host galaxies than
in non-AGN galaxies \citep{arsenault89,knapen+00,laine+02},
while many others found no significant excess of barred galaxies in AGN-host galaxies
\citep{moles95,mcleod95,mulchaey97,ho97,laurikainen+04,hao+09,Ann+09}.

\citet{shlosman89} suggested the ``bars within bars'' scenario as a mechanism for fueling AGNs.
In this model, large-scale stellar bars transport gas into their rotating disks of a few hundred parsec scale.
When a critical amount of gas is accumulated, the disks undergo gravitational instability,
triggering a gaseous secondary bar, which enables gas to approach closer to SMBHs \citep{mulchaey97,maciejewski97,ho97}.
Some studies suggested that nuclear spirals, instead of secondary bars, are responsible for triggering AGNs
\citep{martini99,marquez+00,martini+03}.
In a recent high-resolution smoothed particle hydrodynamics simulation, \citet{hopkins+10} showed that
disk instabilities (for $10-100$ pc scales) driven by primary bars exhibit various morphologies
as well as bar-like shapes.
Several observational studies confirmed the presence of secondary bars by detecting nuclear bars embedded
in large-scale bars, using the ISAAC/VLT spectroscopic data \citep{emsellem+01}, the Hubble Space Telescope (HST)
images \citep{malkan98,laine+02,carollo+02,erwin+02}, and the integral field spectrograph SAURON data \citep{emsellem+06}.
It is found, however, that the fraction of secondary bars in AGNs is similar to that in non-AGNs \citep{martini+03},
implying that secondary bars do not play a critical role in fueling AGNs.
Although there have been many attempts, the nature of the AGN-bar connection is still unclear.
The presence of secondary bars or nuclear spirals in non-AGN galaxies suggests that
bars are not a universal fueling mechanism \citep{marquez+00,laine+02,martini+03}.

In this paper, we investigate the connection between the presence of bars and AGN activity
using a large sample of galaxies from the Sloan Digital Sky Survey (SDSS; \citealt{york+00}).
SDSS data have been used by several previous studies in revealing the dependence of the bar fraction
either on internal galaxy properties or on environmental properties \citep{barazza+08,
aguerri+09,li+09,nair+10b,masters+11,lee+11}.
However, the AGN-bar connection has not been studied in detail using the SDSS data.
\citet{hao+09} found no excess of bars in AGN-host galaxies using SDSS data.
However, the galaxy sample in their study was relatively small and biased to blue galaxies
\citep[see][]{masters+11, lee+11}.
Therefore, it is needed to investigate the connection between bars and AGN activity
using a homogeneous and large galaxy sample.
Following our detailed study on the relation between the presence of bars and
galaxy properties \citep[][hereafter Paper I]{lee+11},
we investigate the AGN-bar connection using a homogeneous sample of late-type galaxies,
selected from the SDSS.

This paper is organized as follows. We describe the volume-limited sample and
the method for identifying bars and spectral types in Section 2.
Section 3 presents the main results including the dependence of the bar fraction on
spectral types, the dependence of the AGN fraction on the presence of bars, and
the comparison of Eddington ratio distributions between barred and non-barred AGN-host galaxies.
We discuss the implication of primary results in Section 4, and present summary and conclusions in Section 5.


\begin{figure*}
\centering
\includegraphics[scale=0.8]{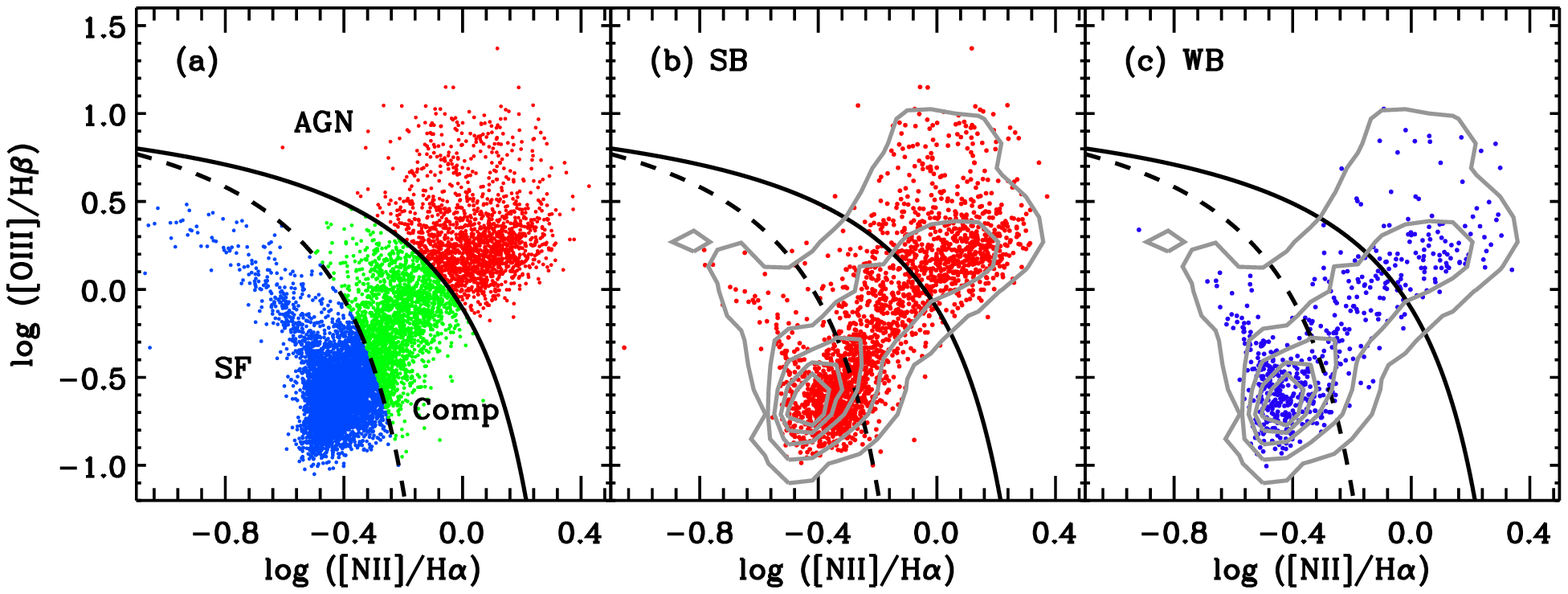}
\caption{(a) Classification of spectral types for 8,655 late-type galaxies in the [NII]/$\ha$ versus [OIII]/$\hb$
diagram. There are three types of galaxies (red: AGN-host galaxies, green: composite galaxies,
and blue: star-forming galaxies) classified by two separate lines, the extreme starburst classification line
(\citealp[solid line]{kewley+01}) and the pure star formation line (\citealp[dashed line]{kauffmann+03}).
In panels (b) and (c), dots represent strong-barred (SB) and weak-barred (WB) galaxies, respectively. Contours show
the distribution for all late-type galaxies shown in panel (a).}
\label{fig1}
\end{figure*}

\section{Data and Methods}

\subsection{SDSS Galaxy Sample}

We use a volume-limited sample of 33,391 galaxies with the $r$-band absolute magnitude
$M_r \leq -19.5+5{\rm log}h$ mag
(hereafter, we drop the $+5{\rm log}h$ term in the absolute magnitude)
at redshift $0.02 \leq z \leq 0.05489$, from the SDSS Data Release 7 (DR7; \citealp{abazajian+09}).
These galaxies are extracted from the Korea Institute for Advanced Study Value-Added Galaxy Catalog
(KIAS VAGC; \citealp{choi+10}) that is based on the Large Scale Structure (LSS) sample of New York
University Value-Added Galaxy Catalog (NYU VAGC; \citealp{blanton+05}).
The rest-frame absolute magnitudes of individual galaxies are computed in fixed bandpass, shifted to $z=0.1$,
using the Galactic reddening correction of \citet{schlegel98} and $K$-corrections as described by \citet{blanton+03}.
The mean evolution correction given by \citet{tegmark+04}, $E(z)=1.6(z-0.1)$, is also applied.
The spectroscopic parameters (i.e., stellar velocity dispersions and strength of various emission lines)
are obtained from NYU VAGC and MPA/JHU DR7 VAGC \citep{tremonti+04,brinchmann+04}.
Stellar masses are also from the MPA/JHU DR7 VAGC, which are based on fits to the SDSS five-band photometry
\citep{kauffmann+03b}.
We adopt a flat $\Lambda$CDM cosmology with $\Omega_{\Lambda}=0.74$ and $\Omega_{m}=0.26$
from {\it Wilkinson Microwave Anisotropy Probe} five-year data \citep{komatsu+09}.

The detailed description of the morphological classification and
the identification of bars of the sample, and the comparison with previous
classifications \citep{rc3,nair+10a} can be found in Paper I (see Section 3).
We summarize the sample selection and classification schemes as follows.
First, after dividing all galaxies into early- and late-type galaxies
using the automated classification method \citep{park+05} and visual inspection,
we selected 19,431 late-type galaxies out of 33,391 galaxies
(see Table 1 of Paper I).
To avoid the internal extinction effects, we selected only galaxies with the minor-to-major axis ratio
$b/a>0.6$, obtaining a sample of 10,674 late-type galaxies.
Then, we classified these late-type galaxies into three groups
based on the presence (and the length) of bars:
2542 strong-barred (23.8\%), 698 weak-barred (6.5\%),
and 7434 non-barred galaxies.
When the size of bars is larger (shorter) than a quarter of the size of their host galaxies,
we classified these galaxies as strong-barred (weak-barred) galaxies.
As described in Paper I, our classification shows a good agreement with
\citet{nair+10a}'s classification.

In this study we classify the sample galaxies into AGNs and non-AGNs using the spectral features.
To determine the spectral types and to investigate
the dependence of the bar fraction on spectral types,
we select galaxies whose spectra show strong emission-lines of $\ha$, $\hb$, $\oth$, and $\ntw$
with signal-to-noise ratio ${\rm S/N}\ge3$ \citep{kewley+06}.
By excluding 2019 galaxies that do not satisfy the S/N criterion,
we make a final the sample of 8655 late-type galaxies for the following analysis.

We perform the aperture correction to the velocity dispersion of the target galaxies
using the equation suggested by \citet{cappellari+06}, \begin{equation}
\sigma_{\rm corr}=\sigma_{\rm fib}\times(R_{\rm fib}/R_{\rm eff})^{(0.066\pm0.035)},
\end{equation}
where $\sigma_{\rm fib}$ is the velocity dispersion obtained from a fiber with $R_{\rm fib}=1^{\prime\prime}.5$.
$R_{\rm eff}$ is an effective radius calculated by $R_{\rm eff}=r_{\rm deV}\times(b/a)^{0.5}_{\rm deV}$
\citep{bernardi+03}, where $r_{\rm deV}$ and $(b/a)_{\rm deV}$ are, respectively,
scale radius and $b/a$ axis ratio in $i$-band in the de Vaucouleurs fit.
Hereafter, the velocity dispersion means $\sigma_{\rm corr}$, and its subscript corr will be omitted.

\subsection{Classification of Spectral Types}


\begin{deluxetable}{crrrrr}
\tablecolumns{6}
\tablewidth{0pc}
\tablecaption{Spectral Types of the Sample Galaxies}
\tablehead{
\colhead{} & \multicolumn{2}{c}{${\rm S/N}\tablenotemark{a}\geq3$} & & \multicolumn{2}{c}{${\rm S/N}\ge6$} \\
\cline{2-3} \cline{5-6}
\colhead{Spectral type} & \colhead{Number} & \colhead{Fraction} & & \colhead{Number} & \colhead{Fraction}}
\startdata
Star-forming & 4,940 & 57.1 \% & & 3,600 & 60.3 \% \\
Composite & 1,973 & 22.8 \% & & 1,411 & 23.7 \% \\
AGN-host & 1,742 & 20.1 \% & & 957 & 16.0 \% \\
\cline{1-6}
Total & 8,655 & 100 \% & & 5,968 & 100 \%
\enddata
\tablenotetext{a}{S/N for four emission lines such as $\ha$, $\hb$, $\ntw$, and $\oth$}
\label{table1}
\end{deluxetable}


\begin{deluxetable*}{crrrrrrr}
\tablecolumns{8}
\tablewidth{0pc}
\tablecaption{Dependence of Bar Fraction on Spectral Types}
\tablehead{
\colhead{(1) ${\rm S/N}\tablenotemark{a}\ge3$} &&&&& \\
\cline{1-8}
\colhead{Spectral type} & \colhead{Total} & \colhead{SB\tablenotemark{b}} & \colhead{$\bfrsbo$ (\%)\tablenotemark{c}} &
\colhead{WB\tablenotemark{d}} & \colhead{$\bfrsbw$ (\%)} & \colhead{SB+WB} & \colhead{$f_{\rm SB+WB}$ (\%)}}
\startdata
Star-forming & 4,940 &  770 & $15.6\pm0.5$ &  348 & $7.0\pm0.4$ & 1,118 & $22.6\pm0.6$ \\
Composite & 1,973    &  639 & $32.4\pm1.1$ &  118 & $6.0\pm0.5$ & 757   & $38.4\pm1.1$ \\
AGN-host     & 1,742 &  742 & $42.6\pm1.1$ &  109 & $6.3\pm0.6$ & 851   & $48.9\pm1.2$ \\
\cline{1-8}
\cline{1-8}
\\
(2) ${\rm S/N}\ge6$ &&&&&& \\
\cline{1-8}
Spectral type & Total & SB & $\bfrsbo$ (\%) &  WB & $\bfrsbw$ (\%) & SB+WB & $f_{\rm SB+WB}$ (\%) \\
\cline{1-8}
Star-forming & 3,600 &  659 & $18.3\pm0.6$ &  265 & $7.4\pm0.4$ & 924   & $25.7\pm0.7$ \\
Composite    & 1,411 &  513 & $36.4\pm1.3$ &   67 & $4.7\pm0.5$ & 580   & $41.1\pm1.3$ \\
AGN-host     & 957   &  419 & $43.8\pm1.6$ &   52 & $5.4\pm0.7$ & 471   & $49.2\pm1.6$
\enddata
\tablenotetext{a}{S/N for four emission lines such as $\ha$, $\hb$, $\ntw$, and $\oth$}
\tablenotetext{b}{Strong-barred galaxies}
\tablenotetext{c}{The errors of the bar fraction are obtained by calculating the standard deviation in
1,000-times-repetitive sampling method.}
\tablenotetext{d}{Weak-barred galaxies}
\label{table2}
\end{deluxetable*}


\begin{figure*}
\centering
\includegraphics[scale=0.8]{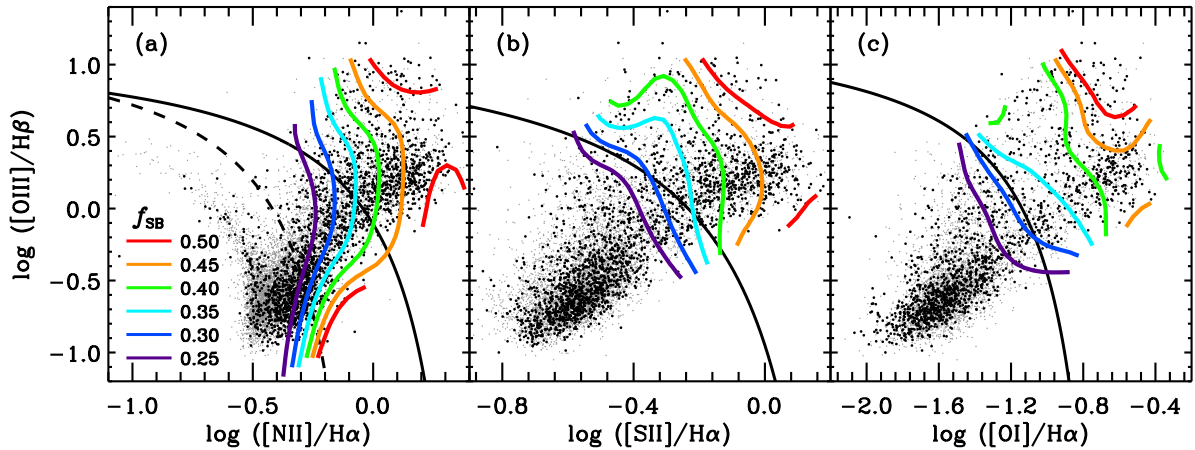}
\caption{The fraction of strong-barred galaxies ($\bfrsbo$) in the three BPT diagnostic diagrams: (a) [NII]/$\ha$ versus [OIII]/$\hb$,
(b) [SII]/$\ha$ versus [OIII]/$\hb$, (c) [NII]/$\ha$ versus [OIII]/$\hb$. Black dots and grey dots
represent strong-barred galaxies and non-SB galaxies, respectively. Contours represent constant $\bfrsbo$.
In panels (b) and (c) we use only 8,508 galaxies with ${\rm S/N_{[SII]}}\geq3$ and 5,622 galaxies with ${\rm S/N_{[OI]}}\geq3$.}
\label{fig2}
\end{figure*}

To determine the spectral types, we use the [NII]/$\ha$ versus [OIII]/$\hb$ diagnostic diagram
that is known as Baldwin-Phillips-Terlevish (BPT) diagram \citep{baldwin81,veilleux87}.
We categorize the sample galaxies into three spectral types: star-forming, composite, and AGN-host galaxies.
As shown in Figure \ref{fig1}, \citet{kewley+01} drew theoretical ``maximum starburst lines''
(solid lines) to define the upper boundary of star-forming galaxies,
and \citet{kauffmann+03} added an empirical demarcation line (dashed line)
to distinguish pure star-forming galaxies from composite galaxies whose spectra are affected
by both star forming nuclei and AGNs.

We perform spectral classification using two criteria of S/N (for $\ha$, $\hb$, [NII], and [OIII]
emission lines): ${\rm S/N}\geq3$ and ${\rm S/N}\geq6$.
When ${\rm S/N}\geq3$ is adopted, we find that the fractions of star-forming, composite, and AGN-host galaxies are
57.1\% (4940 galaxies), 22.8\% (1973 galaxies), and 20.1\% (1742 galaxies), respectively.
On the other hand, in the case of ${\rm S/N}\geq6$, the sample galaxies consist of 60.3\% (3600 galaxies) of star-forming,
23.7\% (1411 galaxies) of composite, and 16.0\% (957 galaxies) of AGN-host galaxies.
The result of spectral classification is summarized in Table \ref{table1}.
As the threshold of S/N increases from 3 to 6, the fraction of AGN-host galaxies decreases 20.1\% to 16.0\%.
This is because a significant fraction of LINERs in the low S/N sample is not included in the high S/N sample,
since LINERs tend to dominate the low S/N sample \citep{cidfernandes+10,cidfernandes+11}.

\section{Results}


\begin{figure*}
\centering
\includegraphics[scale=0.7]{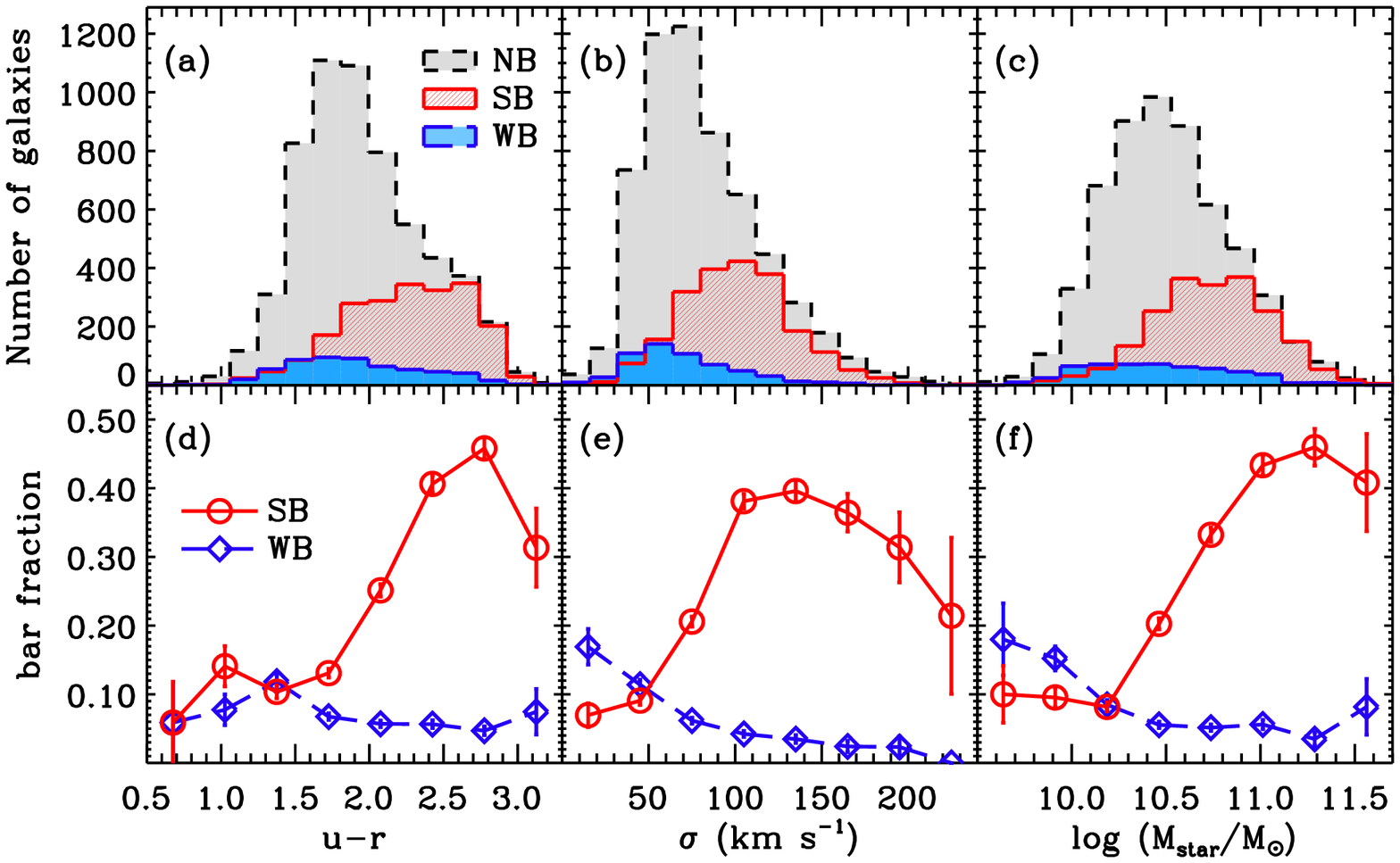}
\caption{(Upper) The number of galaxies as a function of (a) $u-r$ color, (b) velocity dispersion ($\sigma$), and (c) stellar mass ($\mstar$) for
strong-barred (SB), weak-barred (WB), and non-barred (NB) galaxies. (Lower) The dependence of the bar fraction on (d) $u-r$,
(e) $\sigma$, and (f) $\mstar$. Circles and diamonds represent fractions of SB and WB galaxies,
respectively. Error bars mean 1-$\sigma$ sampling errors estimated by calculating the standard deviation of the bar fraction
in 1,000-times-repetitive sampling.}
\label{fig3}
\end{figure*}


\begin{figure}
\centering
\includegraphics[scale=0.55]{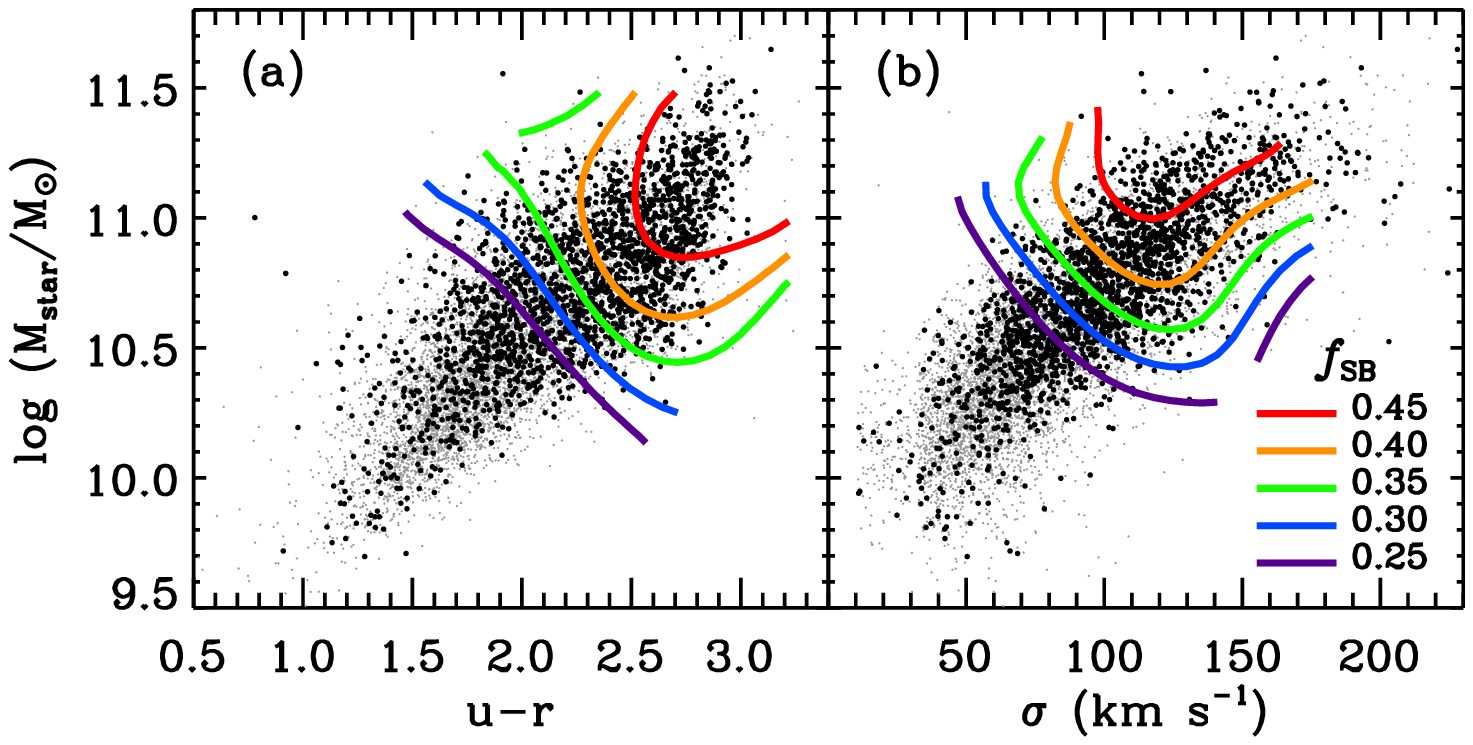}
\caption{The fraction of strong-barred galaxies ($\bfrsbo$) in (a) $u-r$ versus $\mstar$ diagram
and (b) $\sigma$ versus $\mstar$ diagram. Black dots and grey dots represent strong-barred
galaxies and non-SB galaxies, respectively. Contours represent constant strong-barred galaxy fractions.}
\label{fig4}
\end{figure}

To investigate the connection between AGN activity and the presence of bars,
first, we compare the bar fractions in AGN-host and non-AGN galaxies
in Section 3.1.
Then, we examine the AGN fraction between barred and non-barred galaxies in Section 3.2.
Finally, using AGN host galaxies, we investigate whether the Eddington ratio distribution
is different depending on the presence of bars in Section 3.3.

\subsection{Dependence of Bar Fraction on AGN activity}


\begin{figure}
\centering
\includegraphics[scale=0.65]{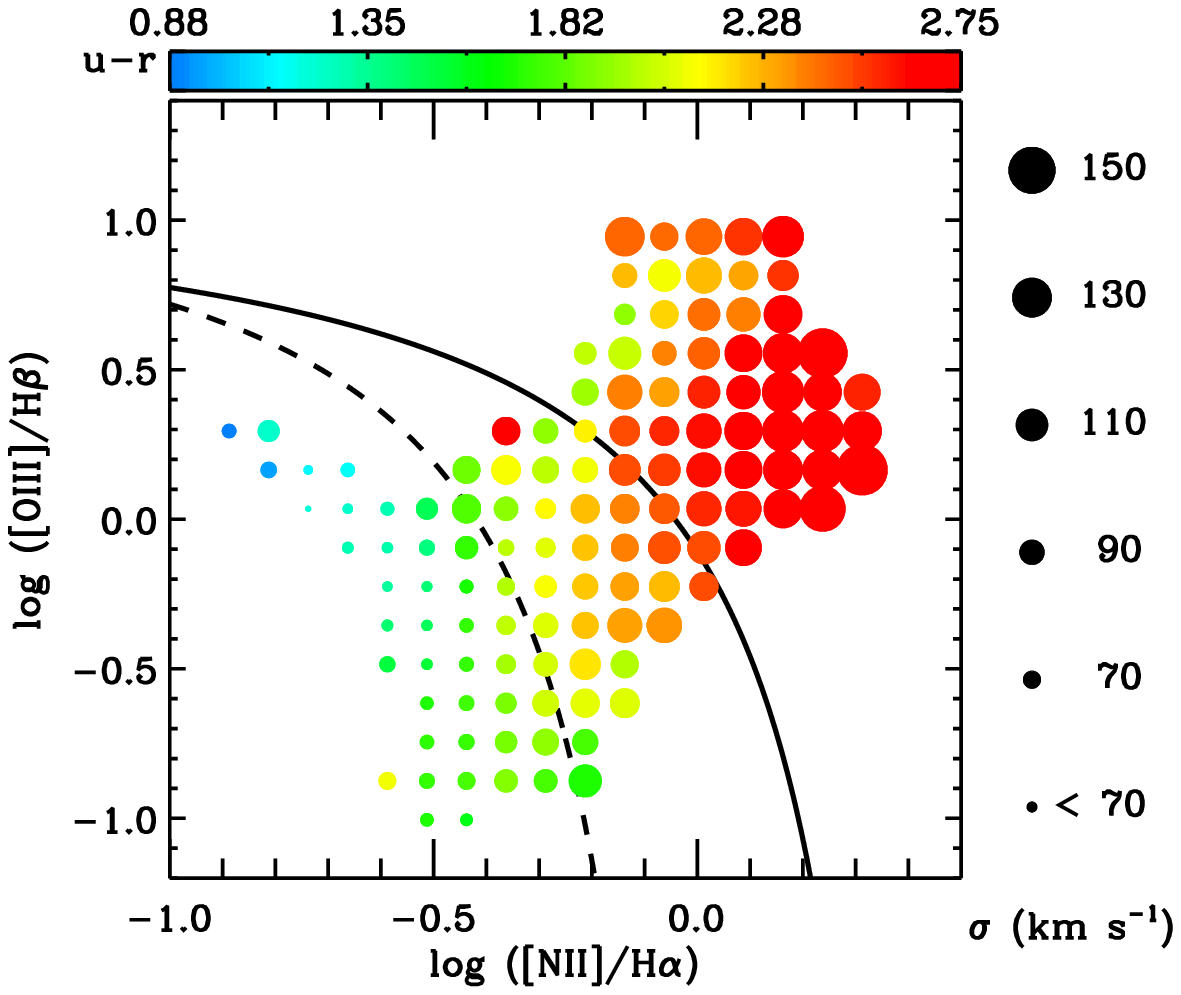}
\caption{Distribution of $u-r$ and $\sigma$ for 8,655 late-type galaxies in [NII]/$\ha$
versus [OIII]/$\hb$ diagnostic diagram. We divide this diagram into $20\times20$ bins and
measure median values of $u-r$ and $\sigma$ for galaxies at each bin.
After excluding bins that contain less than five galaxies, we display
representative symbols with various colors and sizes, corresponding to the median values of $u-r$ and
$\sigma$, respectively.}
\label{fig5}
\end{figure}


\begin{figure*}
\centering
\includegraphics[scale=0.7]{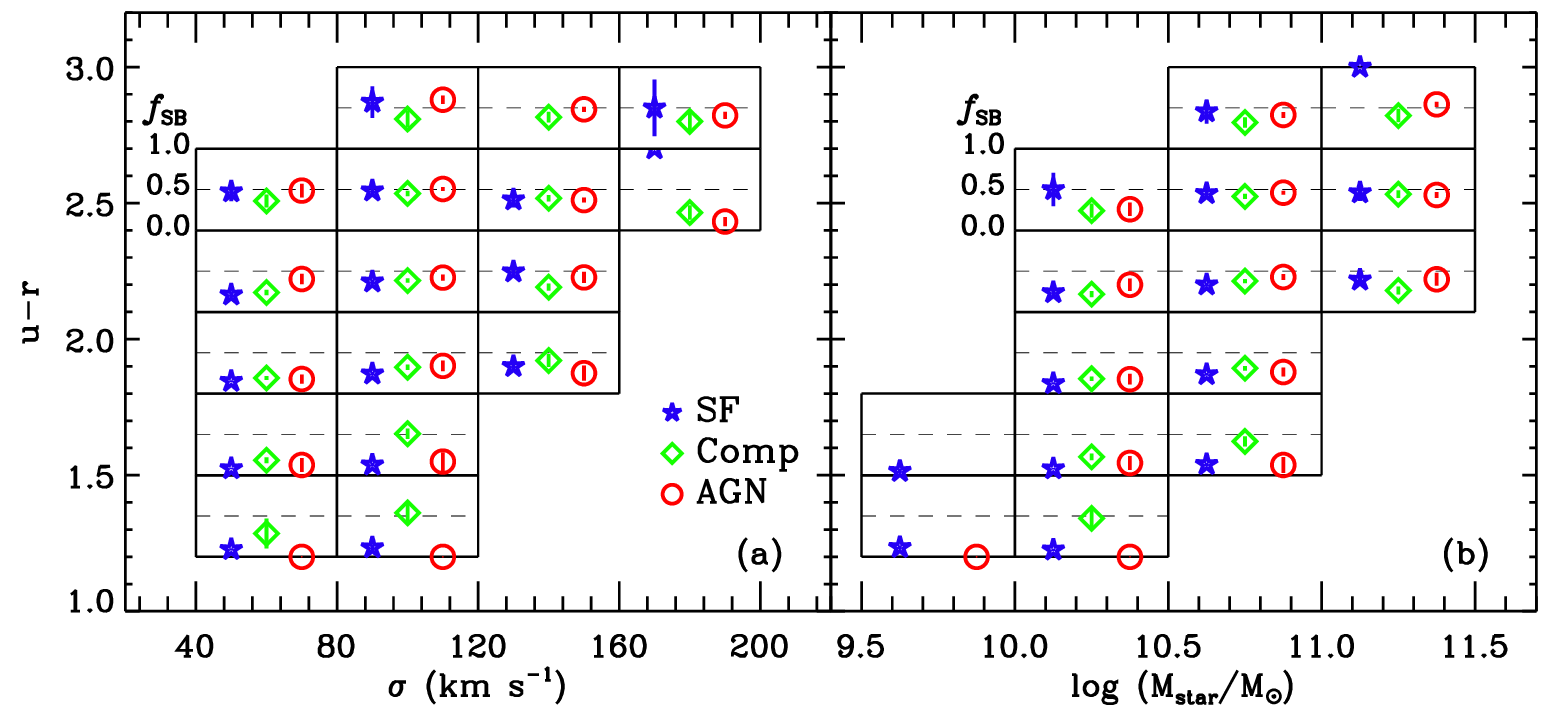}
\caption{The dependence of the fraction of strong-barred galaxies ($\bfrsbo$) on spectral types for late-type galaxies
with fixed ranges of (a) $u-r$ ($\Delta(u-r)=0.3$) and velocity dispersion ($\Delta\sigma=40$ km s${}^{-1}$).
(b) Same as (a), but x-axis is log $(M_{\rm star}/M_{\odot})$ instead of $\sigma$.
Stars, diamonds, and circles with error bars represent $\bfrsbo$ for star-forming, composite,
and AGN-host galaxies, respectively.}
\label{fig6}
\end{figure*}

Figure \ref{fig1} shows the distributions of strong-barred (panel b) and
weak-barred galaxies (panel c) in emission-line ratio diagrams
([NII]/$\ha$ versus [OIII]/$\hb$).
Strong-barred galaxies are widely distributed over the star-forming,
composite, and AGN-host galaxy regions, while the majority of weak-barred galaxies lie
in the star-forming galaxy region.
Some weak bars are found in the composite and AGN-host galaxy region,
but the number of those galaxies is relatively small.

We investigate how the bar fraction varies depending on the spectral types
using two S/N criteria as summarized in Table \ref{table2}.
In the sample of 8655 galaxies with ${\rm S/N}\geq3$, the fraction of strong-barred galaxies ($\bfrsbo$)
is $15.6\%\pm0.5\%$ in star-forming galaxies, $32.4\%\pm1.1\%$ in composite galaxies,
and $42.6\%\pm1.1\%$ in AGN-host galaxies. Among 5968 galaxies with ${\rm S/N}\geq6$, $\bfrsbo$ is
$18.3\%\pm0.6\%$ in star-forming galaxies, $36.4\%\pm1.3\%$ in composite galaxies, and
$43.8\%\pm1.6\%$ in AGN-host galaxies, respectively. In the low S/N case we find that $\bfrsbo$ is $\sim$2.5 times higher
in AGN-host galaxies than in star-forming galaxies. On the other hand, the fraction of weak-barred
galaxies ($\bfrsbw$) does not vary significantly with spectral types, from 6.0\% to 7.0\%.
The result for the high S/N case is not different from that for the low S/N case.

Figure \ref{fig2} presents the change of $\bfrsbo$ for the sample of galaxies with ${\rm S/N}\geq3$ in three BPT diagnostic diagrams
such as (a) [NII]/$\ha$ versus [OIII]/$\hb$, (b) [SII]/$\ha$ versus [OIII]/$\hb$, and (c) [OI]/$\ha$ versus [OIII]/$\hb$.
In panel (b) we use 8508 galaxies with ${\rm S/N_{[SII]}}\geq3$, while only 5622 galaxies with ${\rm S/N_{[OI]}}\geq3$
are used in panel (c).
We find a noticeable trend that $\bfrsbo$ increases continuously from the star-forming galaxy region
(lower left) to the AGN-host galaxy region (upper right) in all diagnostic diagrams,
showing that the presence of bars is more frequent in AGN-host galaxies than non-AGN galaxies.
We check that this trend does not change significantly even when we use the high S/N sample.

In the view of the results we obtained above, it is seen that AGN activity is related to the presence of strong bars.
However, we will see below that these results do not directly indicate a connection between the presence of strong bars and AGN activity.
This is because $\bfrsbo$ is also a strong function of galaxy properties, i.e., $u-r$ color, velocity dispersion ($\sigma$) and stellar mass ($\mstar$).
In Paper I, we found that $u-r$ and $\sigma$ are more influential parameters in determining the bar fraction.
Therefore, we need to compare AGN-host and non-AGN galaxies with fixed $u-r$ and $\sigma$
in order to separate the effect of the two parameters on $\bfrsbo$.

In Figure \ref{fig3} we show the dependence of the bar fraction on three parameters: $u-r$,
$\sigma$, and $\mstar$. $\mstar$ is another important parameter affecting the bar fraction \citep[e.g.,][]{sheth+08,cameron+10,mabreu+10,nair+10b}.
The fraction of strong-barred galaxies increases significantly as $u-r$ color becomes redder,
and it has a maximum value at intermediate velocity dispersion of $\sim$130 km s$^{-1}$.
It has a constant value ($\sim$10\%) until ${\rm log}~(M/M_{\odot})=10.2$, but increases with $\mstar$ thereafter.
On the other hand, the fraction of weak-barred galaxies shows a different dependency on the three parameters.
It has a peak value at a bluer color of $u-r\simeq1.4$, and becomes larger as $\sigma$ or $\mstar$ decreases.
This result on $\mstar$ is consistent with the result of \citet{nair+10b}.

Figure \ref{fig4} shows how $\bfrsbo$ varies in the $u-r$ versus $\mstar$ and in the $\sigma$ versus $\mstar$
diagrams. It shows that $\mstar$ has a strong correlation with both $u-r$ and $\sigma$.
It also shows that $\bfrsbo$ increases with the three parameters.
However, at $\sigma\gtrsim120$ km s$^{-1}$ it decreases with $\sigma$ at a given $\mstar$ bin.
Particularly, contours are neither vertical nor horizontal,
suggesting that $\bfrsbo$ depends on the three parameters simultaneously.
Therefore, we conclude that $u-r$, $\sigma$, and $\mstar$ are all important parameters
in determining the bar fraction.

Because $\bfrsbo$ is strongly correlated with three parameters,
we need to check whether the trend of $\bfrsbo$ in Figure \ref{fig2} is originated from the effect of the three parameters
on $\bfrsbo$. To investigate the difference of $u-r$ and $\sigma$ between AGN-host
and non-AGN galaxies, we examine how $u-r$ and $\sigma$ vary
in the [NII]/$\ha$ versus [OIII]/$\hb$ diagnostic diagram as shown in Figure \ref{fig5}.
We use $20\times20$ bins in this diagram to measure median values of $u-r$ and $\sigma$
for late-type galaxies at each bin.
After excluding bins that contain less than five galaxies,
we display representative symbols with various colors and sizes,
corresponding to the median values of $u-r$ and $\sigma$, respectively.
From the star-forming galaxy region toward the AGN-host galaxy region,
it seems obvious that $u-r$ color becomes redder.
At the same time, $\sigma$ increases along the same direction.
The median values of $u-r$ are $1.77\pm0.01$, $2.20\pm0.01$ and $2.59\pm0.01$,
respectively for star-forming, composite, and AGN-host galaxies.
In the case of $\sigma$, the median values for star-forming, composite, and AGN-host
galaxies are $66.2\pm0.5$, $95.1\pm0.7$, and $118.3\pm0.9$ km s$^{-1}$, respectively.
Considering the dependence of $\bfrsbo$ on $u-r$ and $\sigma$,
it is clear that the trend shown in Figure \ref{fig2} is caused by the fact that
$\bfrsbo$ increases with $u-r$ and $\sigma$ (or $\mstar$).

To remove the effect of $u-r$, $\sigma$, and $\mstar$, we investigate
the bar fraction in AGN-host and non-AGN galaxies at fixed $u-r$ and $\sigma$ (or $\mstar$),
as shown in Figure \ref{fig6}.
First, we divide our sample into 17 bins with fixed $u-r$ and $\sigma$ ranges.
Note that each bin contains more than fifty galaxies.
Then we measure $\bfrsbo$ of each spectral type in each bin.
An obvious excess of $\bfrsbo$ in AGN-host galaxies is found only in one bin with
$u-r=2.1-2.4$ and $\sigma=40-80$ km s$^{-1}$.
In contrast, at all other $u-r$ and $\sigma$ bins, $\bfrsbo$ of AGN-host galaxies is similar
to that of star-forming galaxies.
Second, we perform similar analysis at fixed $u-r$ and $\mstar$ ranges.
We do not found any clear or significant excess of $\bfrsbo$ in AGN-host galaxies in all bins.
These results clearly demonstrate that the $\bfrsbo$ excess in AGN-host galaxies shown in Figure \ref{fig2}
is caused by the fact that on average AGN-host galaxies are redder and and more massive than non-AGN galaxies.
These results suggest that the bar fraction do not depend on AGN activity.

\subsection{Dependence of AGN Fraction on Bar Presence}


\begin{deluxetable*}{crrrrrrrr}
\tablecolumns{9}
\tablewidth{0pc}
\tablecaption{Spectral Classification in Different Bar Types}
\tablehead{
\colhead{(1) ${\rm S/N}\tablenotemark{a}\geq3$} &&&&&&&& \\
\cline{1-9}
\colhead{} & \colhead{NB\tablenotemark{b}} &&& \colhead{SB\tablenotemark{c}} &&& \colhead{WB\tablenotemark{d}} & \\
\cline{2-3} \cline{5-6} \cline{8-9}
\colhead{Spectral type} & \colhead{Number} & \colhead{Fraction} && \colhead{Number} & \colhead{Fraction} &&
\colhead{Number} & \colhead{Fraction}}
\startdata
Star-forming & 3,822 &  64.5\% & &  770  &  35.8\% & & 348 &  60.5\% \\
Composite    & 1,216 &  20.5\% & &  639  &  29.7\% & & 118 &  20.5\% \\
AGN-host     & 891   &  15.0\% & &  742  &  34.5\% & & 109 &  19.0\% \\
\cline{1-9}
Total        & 5,929 & 100.0\% & & 2,151 & 100.0\% & & 575 & 100.0\% \\
\cline{1-9}
\cline{1-9}
\\
(2) ${\rm S/N}\ge6$ &&&&&&&& \\
\cline{1-9}
& ${\rm NB~~~}$ & & & ${\rm SB~~~}$ & & & ${\rm WB~~~}$ & \\
\cline{2-3} \cline{5-6} \cline{8-9}
Spectral type & Number & Fraction && Number & Fraction && Number & Fraction \\
\cline{1-9}
Star-forming & 2,676 &  67.0\% & &  659  &  41.4\% & & 265 &  69.0\% \\
Composite    & 831   &  20.8\% & &  513  &  32.2\% & &  67 &  17.4\% \\
AGN-host     & 486   &  12.2\% & &  419  &  26.3\% & &  52 &  13.5\% \\
\cline{1-9}
Total        & 3,993 & 100.0\% & & 1,591 & 100.0\% & & 384 & 100.0\%
\enddata
\tablenotetext{a}{S/N for four emission lines such as $\ha$, $\hb$, $\ntw$, and $\oth$}
\tablenotetext{b}{Non-barred galaxies}
\tablenotetext{c}{Strong-barred galaxies}
\tablenotetext{d}{Weak-barred galaxies}
\label{table3}
\end{deluxetable*}


\begin{figure}
\centering
\includegraphics[scale=0.5]{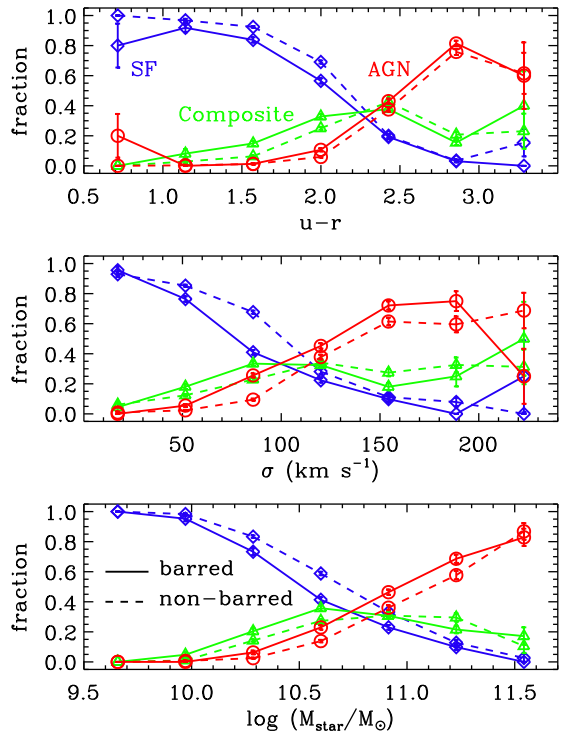}
\caption{The fraction of AGN-host (circles), composite (triangles), and star-forming galaxies (diamonds) as a function
of $u-r$ (top), $\sigma$ (middle), and $\mstar$ (bottom). Solid lines and dashed lines represent barred (strong and weak-barred)
and non-barred galaxies, respectively. Error bars mean 1-$\sigma$ sampling errors.}
\label{frac_agn_mass}
\end{figure}


\begin{figure*}
\centering
\includegraphics[scale=0.65]{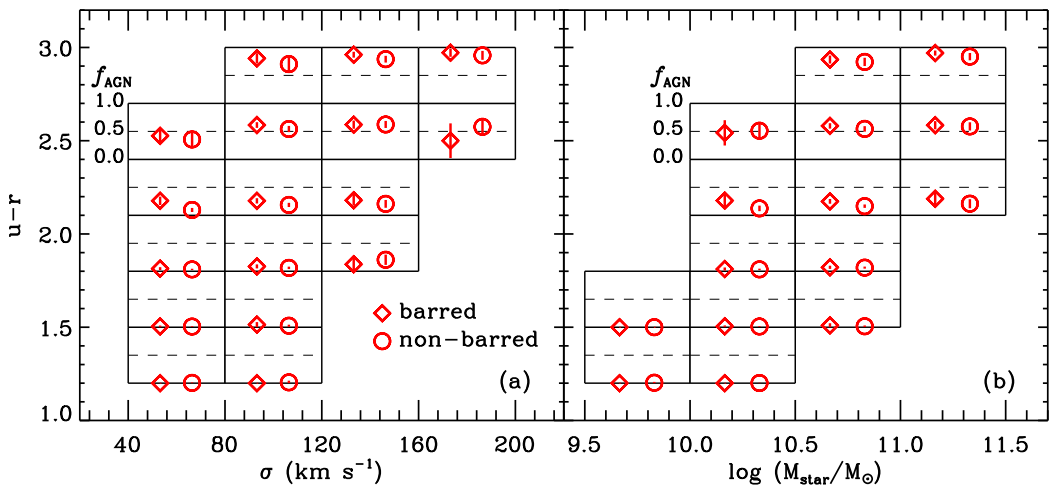}
\caption{The dependence of the AGN fraction ($f_{\rm AGN}$) on the presence of bars at
fixed ranges of (a) $u-r$ and $\sigma$ and (b) $u-r$ and $\mstar$. Diamonds and circles
represent $f_{\rm AGN}$ for barred (strong and weak-barred) and non-barred galaxies, respectively.
Error bars represent 1-$\sigma$ sampling errors.}
\label{agnfrac_barredness}
\end{figure*}

In this section we investigate how AGN fraction changes
depending on the presence of bars.
Among non-barred or weak-barred galaxies,
the fraction of star-forming galaxies ($>60\%$) is much larger than
that of AGN-host galaxies ($<20\%$)
while composite galaxies occupy $\sim$20\%.
In contrast, the AGN fraction increases by a factor of two
when galaxies have strong bars.
For example, among galaxies with ${\rm S/N}\geq3$
(${\rm S/N}\geq6$), the AGN fraction is $34.5\%$ ($26.3\%$).
AGN fractions in strong-barred, weak-barred, and non-barred galaxy samples
are summarized in Table \ref{table3}.

To demonstrate the dependence on galaxy properties,
we present AGN fraction of barred (combining strong and weak bars)
and non-barred galaxies,
as a function of $u-r$, $\sigma$, and $\mstar$ in Figure \ref{frac_agn_mass}.
In both barred and non-barred galaxy samples, the AGN fraction increases
with redder color, higher $\sigma$ or larger $\mstar$
while the fraction of star-forming galaxies shows an opposite trend.
At fixed $\sigma$ or $\mstar$, AGN fraction is significantly higher in barred galaxies
than in non-barred galaxies.
At fixed $u-r$ color, however, the excess of AGN fraction in barred galaxies
is marginal.
These results do not significantly change even if we exclude weak bars
from the barred galaxy sample.
The results presented in Figure \ref{frac_agn_mass} are
similar to the finding of \citet[][see their Figure 8]{oh+11},
which appeared in the literature during the review process of our manuscript.
We note that the sample size in this study is much larger than
that in Oh et al. (8655 galaxies versus 3934 galaxies), and that we classify
composite galaxies separately instead of including them in the AGN sample.

The results in Figures \ref{fig3}, \ref{fig4}, and \ref{frac_agn_mass} show that
both bar fraction and AGN fraction increase as galaxy color becomes redder.
This leads naturally to an expectation that AGN and bars may be related.
However, this does not necessarily mean that both are related. We need to investigate
whether AGN is directly connected to bars or not.

For this we investigate whether AGN fraction is different between barred and non-barred galaxies
in multi-dimensional spaces as shown in Figure \ref{agnfrac_barredness}.
When we compare galaxies at fixed ranges of $u-r$ and $\sigma$ (or $\mstar$),
the excess of AGN fraction in barred galaxies disappears or weakens.
Within the sampling errors,
there is no significant difference of AGN fraction between barred and non-barred galaxies.
These results are dramatically different from those in Figure \ref{frac_agn_mass}
due to the fact that AGN fraction is dependent of color and $\sigma$ (or $\mstar$).
Figure \ref{fig4} shows that $u-r$ has a large dispersion even when $\mstar$ is fixed.
Similarly large color dispersion is also seen when $\sigma$ is fixed (see Figure 9 in Paper I).
Therefore an excess of the AGN fraction in barred galaxies shown in Figure \ref{frac_agn_mass} is an
``apparent'' trend, which is caused by the residual dependence on $u-r$ color in $\sigma$
(or $\mstar$) bins.
Thus, in order to remove the effect of $u-r$ and $\sigma$ (or $\mstar$),
AGN fraction has to be compared using galaxies at fixed $u-r$ {\it AND} $\sigma$ (or $\mstar$).
We find no significant dependence of AGN fraction on the presence of bars,
suggesting that AGN activity is not dominated by bars.


\begin{figure*}
\centering
\includegraphics[scale=0.8]{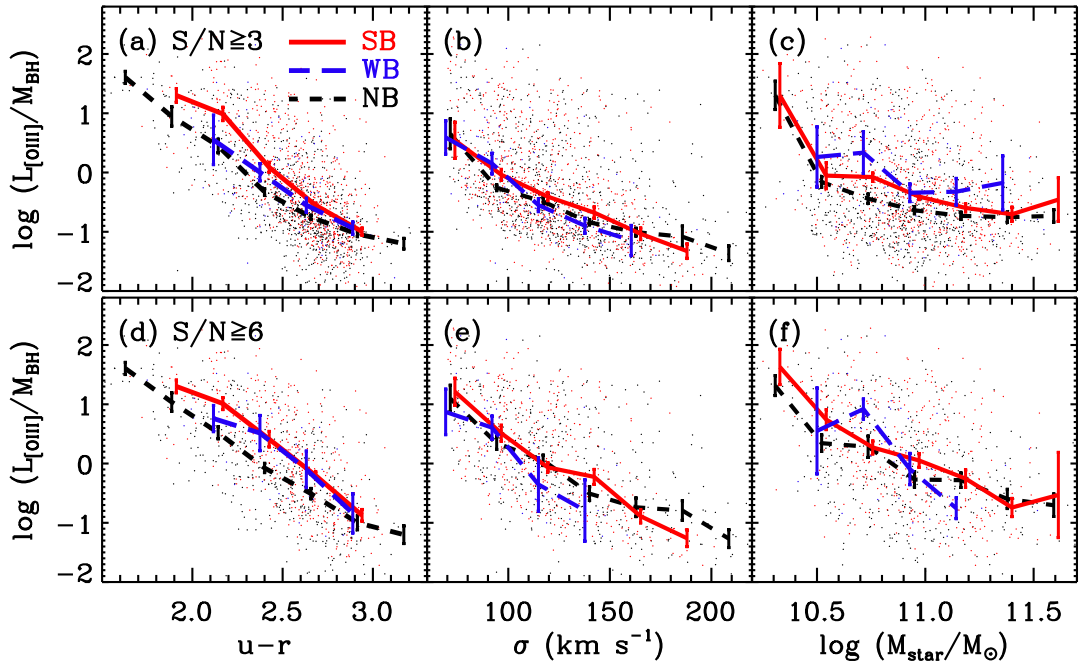}
\caption{Eddington ratio as a function of (left) $u-r$, (middle) $\sigma$,
and (right) $\mstar$ for 1,647 (upper, ${\rm S/N}\geq3$) and 893 (lower, ${\rm S/N}\geq6$)
late-type AGN-host galaxies with $\sigma>70$ km s${}^{-1}$. Solid lines, long-dashed lines, and short-dashed lines represent
median curves of Eddington ratio for strong-barred (SB), weak-barred (WB), and non-barred (NB) AGN-host galaxies,
respectively. The size of bins corresponds to 1/7 of x-axis of each panel, and only when bins contain more than
five galaxies median values of Eddington ratio are drawn. Error bars are calculated using 1,000-times resampling
method. Error bars for strong-barred and weak-barred galaxies are slightly shifted respect to ones of non-barred galaxies
in order to avoid overlap each other.}
\label{fig7}
\end{figure*}


\begin{figure}
\centering
\includegraphics[scale=0.6]{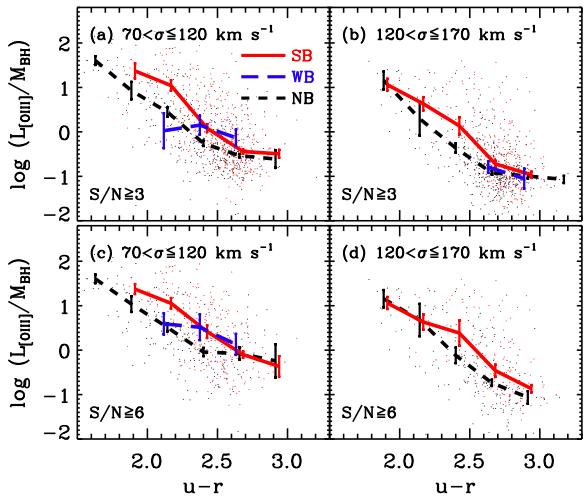}
\caption{Eddington ratio versus $u-r$ for late-type AGN-host galaxies with (upper) ${\rm S/N}\geq3$, (lower)
${\rm S/N}\geq6$, (left) 70 km s$^{-1}<\sigma\leq120$ km s$^{-1}$, and (right) 120 km s$^{-1}<\sigma\leq170$ km s$^{-1}$.
Solid lines, long-dashed lines, and short-dashed lines represent median curves of
Eddington ratio for strong-barred (SB), weak-barred (WB), and non-barred (NB) galaxies, respectively.
The size of bins corresponds to 1/7 of x-axis of each panel, and only when bins contain
more than five galaxies median values of Eddington ratio are drawn. Error bars are calculated using 1,000-times resampling
method. Error bars for strong-barred and weak-barred galaxies are slightly shifted respect to ones of non-barred galaxies
in order to avoid overlap each other.}
\label{fig8}
\end{figure}

\subsection{Comparison of Eddington Ratio between Barred and Non-barred AGN-host Galaxies}

If AGN activity is triggered by bars, barred galaxies may have
higher accretion rates than non-barred galaxies.
We exclusively use AGN-host galaxies to examine whether there is any difference
in AGN power between barred and non-barred galaxies.

We use the [OIII] luminosity ($L_{\rm [OIII]}$) as a proxy for the bolometric luminosity
and infer black hole mass ($M_{\rm BH}$) from $\sigma$ using Eq. (2) as described below.
Thus, $L_{\rm [OIII]}$ to $M_{\rm BH}$ ratio can be used as an approximate
Eddington ratio indicator.
We adopt a reddening curve of $R_{V}=A_{V}/E(B-V)=3.1$ \citep{cardelli89}
and an intrinsic Balmer decrement of $\ha/\hb=3.1$ for AGN-host galaxies \citep{osterbrock+06}
to correct for internal dust extinction.
To estimate $M_{\rm BH}$, we adopt a $M_{\rm BH}-\sigma$ relation
for late-type galaxies suggested by \citet{mcconnell+11}:
\begin{equation}
{\rm log}~(M_{\rm BH}/M_{\odot}) = 7.97+4.58\times {\rm log}~(\sigma/200 {\rm km ~s^{-1}}).
\end{equation}

We exclude galaxies with velocity dispersion values lower than the instrumental
resolution of the SDSS spectra ($\sigma<70$ km s${}^{-1}$) since these measurements
are not reliable \citep{choi+09}.
In the following analysis we use two samples: 1647 AGN-host galaxies with ${\rm S/N}\geq3$ and 893 ones with ${\rm S/N}\geq6$.
The low S/N sample contains 718 strong-barred, 97 weak-barred, and 832 non-barred galaxies. On the other hand,
the high S/N sample consists of 407 strong-barred, 46 weak-barred, and 440 non-barred galaxies.
The black hole mass of AGN-host galaxies spans $5.9 <$ log ($M_{\rm BH}/M_{\odot}$)$ < 8.3$,
while $\eddr$ ranges over four order of magnitude,
$10^{-2}-10^{2}$ $L_{\odot}/M_{\odot}$.

In Figure \ref{fig7} we present the Eddington ratio indicator (hereafter Eddington ratio) distributions
as a function of $u-r$, $\sigma$, and $\mstar$, respectively,
for strong-barred, weak-barred, and non-barred galaxies.
The AGN power (i.e., Eddington ratio) appears to decrease with $\mstar$ (or $\sigma$)
as previously seen by \citet{hwang+11}.
This trend can be interpreted as ``Eddington incompleteness'', which reflects the observational
selection effect that for given flux (or luminosity limits) lower Eddington ratio AGNs can be detected at higher mass scales.
Since at a fixed Eddington ratio it is harder to detect [OIII] lines for lower mass black holes,
the Eddington ratios can be distributed down to a much lower values
for higher mass black holes and galaxies as shown in Figure \ref{fig7}.
In addition, we find that the AGN power is correlated with $u-r$ of galaxies.
Blue AGN-host galaxies show significantly higher Eddington ratio than red AGN-host
galaxies, implying that gas-rich systems generally have higher Eddington ratio
than gas-poor systems. However, we note that this correlation can be also caused by
the Eddington incompleteness since bluer color galaxies have on average lower galaxy (hence black
hole) mass.

Nevertheless, we can compare the distributions of the Eddington ratio
among strong-barred, weak-barred, and non-barred galaxies.
We measure median values of log ($\eddr$) as a function of $u-r$, $\sigma$, and $\mstar$
for each bar class.
In panel (a) and (d), it is shown that median curves for strong-barred and weak-barred galaxies
lie slightly above those for non-barred galaxies,
although the differences between barred and non-barred
galaxies are not significant and the median Eddington ratios are consistent
within the error.
In other panels there is no difference between barred
and non-barred AGN-host galaxies over the whole ranges of $\sigma$ and $\mstar$.

To avoid the effect of Eddington incompleteness, we plot the Eddington ratio versus $u-r$ diagram
at two fixed $\sigma$ ranges in Figure \ref{fig8}.
The anti-correlation between the Eddington ratio and $u-r$ is still
present for both 70 km s$^{-1}<\sigma\le120$ km s$^{-1}$ and 120 km s$^{-1}<\sigma\le170$ km s$^{-1}$ ranges,
indicating that AGN power is correlated with the amount of cold gas in galaxies.
\citet{choi+09} also found a similar result among late-type AGN-host galaxies with $7<{\rm log}$
$M_{\rm BH}/M_{\odot}<8$.
We note that the contributions to the [OIII] lines from star formation can
systematically increase the $L_{\rm [OIII]}$/$M_{\rm BH}$ ratio in bluer galaxies.
Thus, further analysis is required to explore the connection between
the presence of gas in large scales and the Eddington ratio.

When we compare the Eddington ratio distributions of barred and non-barred
galaxies in Figure \ref{fig8}, the median values of the Eddington ratio are not significantly
different, implying that AGN power is not strongly affected by the presence of bars.
Considering the scatter in each bin, and the uncertainty and systematic errors in estimating
black hole masses from the $M_{\rm BH}-\sigma$ relation, we conclude that
there is no strong difference of the Eddington ratios between barred and non-barred
galaxies.

\section{Discussion}


\begin{figure*}
\centering
\includegraphics[scale=0.8]{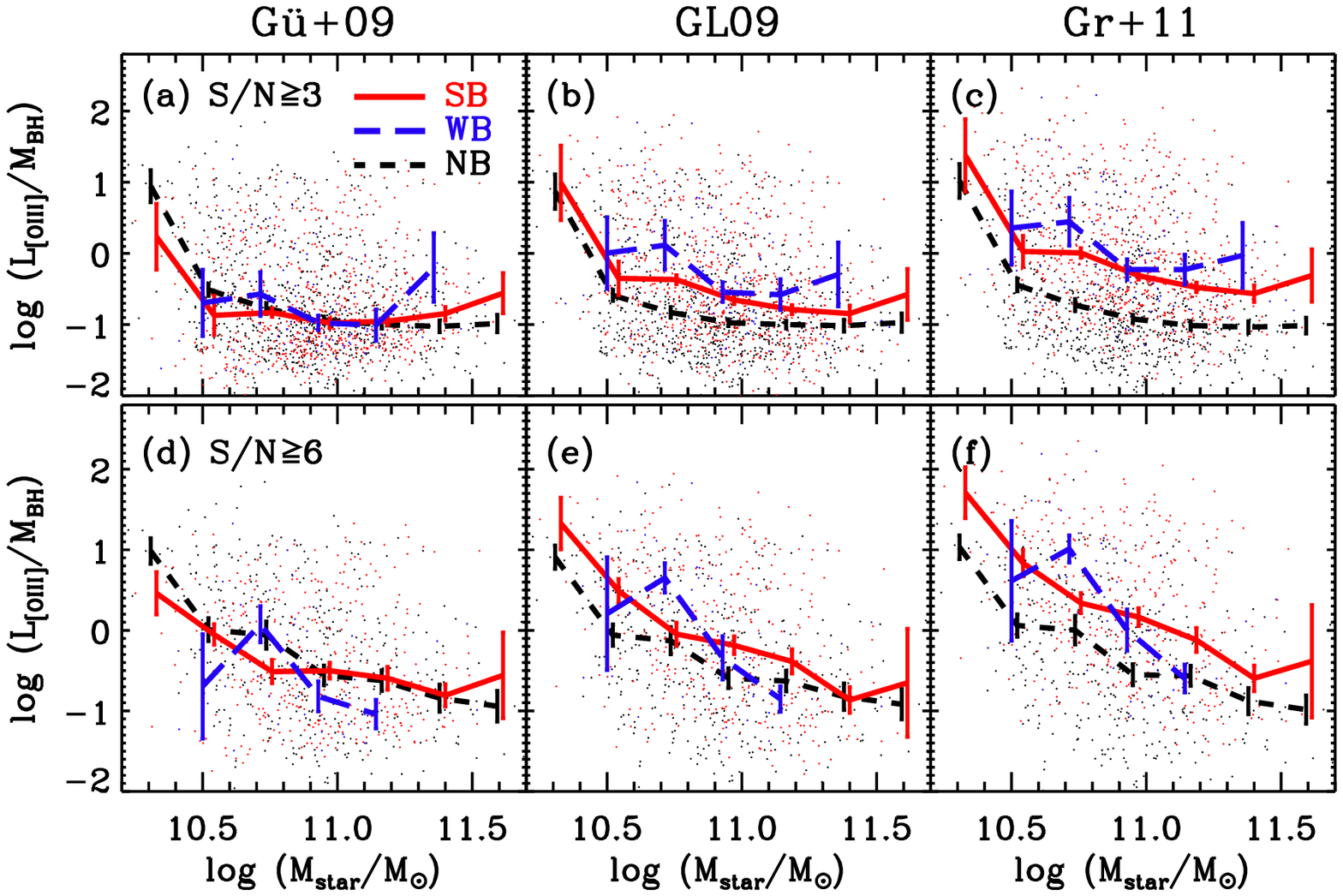}
\caption{Comparison of Eddington ratio as a function of $\mstar$
between strong-barred (SB), weak-barred (WB), and non-barred (NB) AGN-host galaxies with (upper)
${\rm S/N}\geq3$, (lower) ${\rm S/N}\geq6$,
when adopting $M_{\rm BH}-\sigma$ relations given by (left) \citet{gultekin+09}: $(\alpha,\beta)=(7.67,1.08)$ for barred and
(8.19,4.21) for non-barred, (middle) \citet{graham+09}: (8.03,3.94) for barred and (8.15,3.89) for non-barred, and
(right) \citet{graham+11}: (7.80,4.34) for barred and (8.25,4.57) for non-barred galaxies, respectively.}
\label{comp_Eddr_diffmsigrel}
\end{figure*}

\subsection{Do AGNs Favor Barred Galaxies?}

Over the last three decades, bars have been invoked as a mechanism for fueling SMBHs.
Although some studies claimed that AGNs are more frequently found
in barred galaxies \citep{arsenault89,knapen+00,laine+02},
no excess of bars in AGN-host galaxies has been reported by many other statistical studies
\citep{moles95,mcleod95,mulchaey97,ho97,laurikainen+04,hao+09}.
This discrepancy was at least in part caused by the small sample size, selection effect,
and contamination owing to the correlation between bars and other galaxy properties,
i.e., color and stellar mass.

In this study we find that
the fraction of strong bars is $\sim$2.5 times higher in AGN-host galaxies than
in non-AGN galaxies. This result is clearly different from the findings by \citet{hao+09},
who claimed no excess of bar fraction in AGN-host galaxies
based on a sample of 1,144 SDSS disk galaxies with $-18.5>M_{g}>-22.0$ at $0.01<z<0.03$.
The discrepancy is due to the combination of two effects.
First, by using a color cut \citep{bell+04,barazza+08} in selecting disk galaxies,
\citet{hao+09} inevitably excluded red disk galaxies,
leading to a much lower AGN fraction in their sample (11.1\%) than
that in our sample (17.0\%).
Second, they used the ellipse fitting method to identify bars.
In general the bar fraction based on the ellipse fitting \citep[e.g., $\sim$50\% in $r$-band:][]{barazza+08} is much higher than that
obtained by visual inspection (e.g., $\sim$33\% in $B$-band: \citealt{rc3}, $26\%\pm0.5\%$ in $g+r+i$ color images:
\citealt{nair+10a}, $29.4\%\pm0.5\%$ from Galaxy Zoo: \citealp{masters+10}, 30.4\%: Paper I, $36\%$: \citealt{oh+11}).
Thus, the combined effects of differences in sample selection and the
classification method between Hao et al. and ours result in different findings.

We also find that the AGN fraction is twice higher in strong-barred galaxies than in non-barred galaxies
as previous studies similarly reported \citep{arsenault89,knapen+00,laine+02}.
As shown in Figure \ref{frac_agn_mass},
the higher AGN fraction in barred systems is present over a large ranges of
$\mstar$. This result is consistent with those of recent studies \citep{coelho+11,oh+11}.

However, the excess of the bar fraction in AGN-host galaxies and
the excess of the AGN fraction in barred galaxies do not indicate
that the presence of bars and AGN activity are directly connected
since the excess of the bar fraction in AGN-host galaxies and
the excess of the AGN fraction in barred galaxies disappear
when we compare galaxies with the same $u-r$ and $\sigma$ (or $\mstar$)
(see Figure \ref{fig6} and Figure \ref{agnfrac_barredness}).
Thus, we conclude that AGN activity is not dominated by the presence of bars.

Comparing the Eddington ratios among AGN-host galaxies,
we find no significant difference between barred and
non-barred galaxies at fixed $u-r$, $\sigma$ and $\mstar$ bins
(see Figure \ref{fig7}),
indicating that AGN power is not enhanced by the presence of bars.

Among AGN host galaxies with $2.0<u-r<2.5$,
the Eddington ratios are marginally higher in strong-barred
galaxies than in non-barred galaxies (see Figure \ref{fig8}),
possibly implying that strong bars can boost AGN activity when their host galaxies
lie in green valley.
However, the number of AGN-host galaxies in our sample is not enough
to draw a clear conclusion.

\subsection{Dependence on the $M_{\rm BH}-\sigma$ Relation}

Since we estimate black hole masses from stellar velocity dispersion utilizing
the $M_{\rm BH}-\sigma$ relation, the derived Eddington ratios depend on the slope
and intercept of the $M_{\rm BH}-\sigma$ relation. Thus, it is necessary
to investigate whether the Eddington ratio difference
between barred and non-barred galaxies depends on the adopted
$M_{\rm BH}-\sigma$ relation.

Over the last decade, empirical scaling relations between $M_{\rm BH}$ and $\sigma$ have been
improved as the number of galaxies with $M_{\rm BH}$ measurements
increased \citep[e.g.,][]{ferrarese+00,gebhardt+00,tremaine+02,gultekin+09,graham+09,woo+10}.
By combining two new $M_{\rm BH}$ measurements
with the literature data published before August 2011,
a recent study by \citet{mcconnell+11} provided two $M_{\rm BH}-\sigma$ relations;
${\rm log}~(M_{\rm BH}/M_{\odot}) = \alpha+\beta~{\rm log}~(\sigma/200 {\rm km ~s^{-1}})$
with $(\alpha,\beta) = (8.38\pm0.06,4.53\pm0.40)$ for elliptical/S0 galaxies and
$(7.97\pm0.22,4.58\pm1.25)$ for spiral galaxies.
In this study we adopt the second equation for estimating $M_{\rm BH}$
since our sample is composed of late-type galaxies.

A few studies separately derived $M_{\rm BH}-\sigma$ relations for barred and non-barred galaxies
\citep{gultekin+09,graham+09,graham+11}.
For example, \citet{gultekin+09} reported $(\alpha,\beta) = (8.19\pm0.087,4.21\pm0.446)$
for non-barred and $(7.67\pm0.115,1.08\pm0.751)$ for barred galaxies
while \citet{graham+09} derived $(8.15\pm0.05,3.89\pm0.18)$ for non-barred galaxies, and
$(8.03\pm0.05,3.94\pm0.19)$ for the combined sample of barred and non-barred galaxies.
\citet{graham+11} reported steeper $M_{\rm BH}-\sigma$ relations
with $(8.25\pm0.06,4.57\pm0.35)$ for non-barred galaxies and
$(7.80\pm0.10,4.34\pm0.56)$ for barred galaxies.

To demonstrate the dependence of Eddington ratios on the adopted $M_{\rm BH}-\sigma$ relation,
we compare in Figure \ref{comp_Eddr_diffmsigrel}
Eddington ratios between barred and non-barred galaxies
using three different pairs of $M_{\rm BH}-\sigma$ relations mentioned above.
When the $M_{\rm BH}-\sigma$ relations from \citet{gultekin+09} are used,
barred and non-barred galaxies show no significant difference in Eddington ratios
as similarly shown in Figure \ref{fig7}.
In contrast, Eddington ratios tend to be higher in barred galaxies than in non-barred galaxies
when the $M_{\rm BH}-\sigma$ relations are taken from \citet{graham+09}.
The enhanced Eddington ratios in barred galaxies is particularly noticeable
when using \citet{graham+11}'s $M_{\rm BH}-\sigma$ relations. since $M_{\rm}$
in barred galaxies are significantly reduced due to the lower intercept
of the $M_{\rm BH}-\sigma$ relation.

If we adopt different $M_{\rm BH}-\sigma$ relations respectively for barred and non-barred galaxies,
AGN power appears to be enhanced by bars.
\citet{oh+11} used the $M_{\rm BH}-\sigma$ relations taken from \citet{graham+09}, and
they argued that AGN strength is enhanced by the presence bars.
However, there are several limitations in adopting two different relations
for barred and non-barred.
First, the $M_{\rm BH}-\sigma$ relation of barred galaxies is not well defined
since the relation has been derived with a small number of barred galaxies.
For example, \citet{gultekin+09} used only 8 measurements (and 11 upper limits of $M_{\rm BH}$)
of barred galaxies while
\citet{graham+11} also used only 20 barred galaxies. Second, the $M_{\rm BH}-\sigma$ relation of
non-barred galaxies
are biased to early-type galaxies since early-type galaxies are dominant in the sample.
Third, dynamical mass measurements of black holes in barred galaxies
are much more uncertain
since no stellar dynamical model truly accounts for stellar bars \citep{gultekin+09a}.
Therefore, we decide to use only one $M_{\rm BH}-\sigma$ relation for late-type galaxies
as given in \citet{mcconnell+11}, in order to avoid any systematic uncertainties of the
$M_{\rm BH}-\sigma$ relations between barred and non-barred galaxies.
It is necessary to investigate whether barred galaxies have higher Eddington ratios than non-barred
galaxies when more robust $M_{\rm BH}-\sigma$ relations of barred and non-barred late-type
galaxies become available in the future.

\subsection{What Triggers AGNs?}

Based on the statistical analysis using a large sample of $\sim$9000 late-type galaxies, we
find that AGN activity is not dominated by the presence of bars.
Then, what triggers AGNs?

Several numerical simulations suggested that interactions and mergers between galaxies are main triggers for AGN activity
\citep{noguchi87,hernquist89,barnes91,barnes92,mihos96,dimatteo+05,hopkins+06,debuhr+11}.
This scenario is supported by several observational studies. For example, \citet{sanders88}
showed that ultra-luminous infrared galaxies (ULIRG) and quasars are formed
through the strong interaction or merger between gas-rich spirals.
\citet{bahcall97} found that twenty nearby luminous quasars ($z<0.3$)
in their sample have galaxy companions that are closer than 25 kpc.
In the case of lower luminosity AGNs, i.e., Seyfert galaxies,
minor mergers between gas rich galaxies and with their satellite galaxies
are proposed as a mechanism for triggering AGN activity \citep[e.g.,][]{derobertis98}.
However, some observational studies showed conflicting results. \citet{fuentes88} found
a marginal evidence that Seyfert galaxies interact with their companions that have comparable sizes.
In addition, by investigating the environmental dependence of AGN fraction using the SDSS sample,
\citet{miller+03} claimed that the fraction of AGN-host galaxies is independent on environment
\citep[see also,][]{coziol98,shimada+00,schmitt+01}.

Recently, \citet{martinez+10} reported that AGN-host galaxies (45\%) are more frequent than composite (23\%)
or star-forming galaxies (32\%) in the Hickson compact group environment where galaxy-galaxy interactions occur violently.
\citet{hwang+11} and Choi et al. (in prep.) also found, using the SDSS data, that the AGN fraction increases
as the distance to a nearest late-type neighbor galaxy decreases in both cluster and field environments,
and concluded that AGN activity can be triggered through mergers and interactions between galaxies
when gas supply for AGN is available.
In contrast, \citet{dasilva+11} claimed that merging galaxies with signatures of recent
starburst, found in the green valley, had no detectable AGN activity.
Although it is theoretically clear that mergers and interactions between galaxies
can provide advantages for AGN activity by reducing angular momentum of interstellar medium
and by generating gas inflows to the center of galaxies, observational studies do not
clearly show the connection between galaxy interaction and AGN activity.

Additional mechanisms are expected to occur at subkiloparsec scales
in order to influence the central black hole directly.
Secondary bars residing in the nuclear region, which are
often called as nested bars, nuclear bars or inner bars,
have been a strong candidate. Some observational works
\citep{shaw95,wozniak95,friedli96,mulchaey97,jungwiert97,greusard+00,
emsellem+01,emsellem+06,laurikainen+07} found secondary bars in the central region of galaxies.
These secondary bars are predicted by the ``bars within bars'' scenario
\citep{shlosman89}. The dynamical and kinematical properties of secondary bars are investigated
by simulations \citep{englmaier+04,maciejewski+08,shen+09,shen+11} and by observations \citep{garcia98,schinnerer+06,delorenzo+08}.
Similarly, nuclear dust spirals are invoked as a means to transport material from kiloparsec scales
down to sub-kpc scales. High resolution images from HST provided a close look at dusty structures in
nuclear regions \citep{malkan98,regan99,martini99,pogge+02}.
\citet{martini+03} classified nuclear spirals into several morphological types:
grand-design, tightly wound, loosely wound, chaotic nuclear spirals. Recently, \citet{hopkins+10}
showed that gas structures formed by gravitational instabilities at several parsec
scale have diverse morphologies: spirals, rings, clumps and also bars.
So, they proposed ``stuff within stuff'' model that is a revised version of \citet{shlosman89}'s
model. However, there is another argument that the presence of nuclear spirals is not also directly connected
with current AGN activity, because the frequency of nuclear spirals in AGN-host galaxies is comparable with that in
non-AGN galaxies \citep{martini+03}. Thus, further studies are needed to investigate any relation between
nuclear spirals (or secondary bars) and activity in galactic nuclei.

\section{Summary and Conclusions}

We investigate the relation between the presence of bars and AGN activity,
using a sample of 8655 late-type galaxies with $b/a>0.6$ and $M_{r} < -19.5$
at $0.02\le z\le 0.05489$, selected from the SDSS DR7.
We divide these galaxies into three spectral types: star-forming, composite,
and AGN-host galaxies, and classified them into barred (storng \& weak) and non-barred galaxies
by visual inspection.
We summarize our main findings as follows.

1. The strong bar fraction is $\sim$2.5 times higher in AGN-host galaxies
than in star-forming galaxies. However, the excess of $\bfrsbo$
is caused by the fact that AGN-host galaxies have on average redder $u-r$ color
and higher $\sigma$ than non-AGN galaxies
since $\bfrsbo$ is higher for redder and more massive
(higher $\sigma$) galaxies.
The excess of $\bfrsbo$ in AGN-host galaxies disappears when
galaxies with the same $u-r$ and $\sigma$ (or $\mstar$) are compared,
indicating that $\bfrsbo$ do not depend on AGN activity.

2. Strong-barred galaxies have higher AGN fraction than weak-barred
or non-barred galaxies.
However, we find no difference of the AGN fraction between barred and non-barred galaxies,
when we compare galaxies with the same $u-r$ color and $\sigma$ (or $\mstar$),
indicating that AGN activity is not dominated by the presence of bars.

3. Among AGN-host galaxies, barred and non-barred systems
show similar Eddington ratio distributions as a function of $u-r$,
$\sigma$, and $\mstar$, implying that AGN power is not enhanced by bars.

In conclusion we do not find any evidence that bars trigger AGN activity.
Thus we argue that there is no direct connection between AGN activity and the presence of bars.

\vspace{1cm}

We thank the anonymous referee for his/her useful comments which improved significantly the original manuscript.
G.H.L. thank Changbom Park and Yun-Young Choi for providing help in producing the KIAS VAGC and performing
the morphology classification.
M.G.L. was supported in part by Mid-career Research Program through the NRF grant funded by the MEST (no.2010-0013875).
J.H.W. acknowledges support by the Basic Science Research Program through the NRF funded by the MEST (no.2010-0021558).
H.S.H. acknowledges the Centre National d'Etudes Spatiales (CNES) and the Smithsonian Institution for the support of
his post-doctoral fellowship.

\end{document}